\documentclass{kapproc} 

\newread\epsffilein    
\newif\ifepsffileok    
\newif\ifepsfbbfound   
\newif\ifepsfverbose   
\newdimen\epsfxsize    
\newdimen\epsfysize    
\newdimen\epsftsize    
\newdimen\epsfrsize    
\newdimen\epsftmp      
\newdimen\pspoints     
\def\epsfbox#1{\global\def\epsfllx{72}\global\def\epsflly{72}%
   \global\def\epsfurx{540}\global\def\epsfury{720}%
   \def\lbracket{[}\def\testit{#1}\ifx\testit\lbracket
   \let\next=\epsfgetlitbb\else\let\next=\epsfnormal\fi\next{#1}}%
\def\epsfgetlitbb#1#2 #3 #4 #5]#6{\epsfgrab #2 #3 #4 #5 .\\%
   \epsfsetgraph{#6}}%
\def\epsfnormal#1{\epsfgetbb{#1}\epsfsetgraph{#1}}%
\def\epsfgetbb#1{%
\openin\epsffilein=#1
\ifeof\epsffilein\errmessage{I couldn't open #1, will ignore it}\else
   {\epsffileoktrue \chardef\other=12
    \def\do##1{\catcode`##1=\other}\dospecials \catcode`\ =10
    \loop
       \read\epsffilein to \epsffileline
       \ifeof\epsffilein\epsffileokfalse\else
          \expandafter\epsfaux\epsffileline:. \\%
       \fi
   \ifepsffileok\repeat
   \ifepsfbbfound\else
    \ifepsfverbose\message{No bounding box comment in #1; using defaults}\fi\fi
   }\closein\epsffilein\fi}%
\def\epsfsetgraph#1{%
   \epsfrsize=\epsfury\pspoints
   \advance\epsfrsize by-\epsflly\pspoints
   \epsftsize=\epsfurx\pspoints
   \advance\epsftsize by-\epsfllx\pspoints
   \epsfxsize\epsfsize\epsftsize\epsfrsize
   \ifnum\epsfxsize=0 \ifnum\epsfysize=0
      \epsfxsize=\epsftsize \epsfysize=\epsfrsize
     \else\epsftmp=\epsftsize \divide\epsftmp\epsfrsize
       \epsfxsize=\epsfysize \multiply\epsfxsize\epsftmp
       \multiply\epsftmp\epsfrsize \advance\epsftsize-\epsftmp
       \epsftmp=\epsfysize
       \loop \advance\epsftsize\epsftsize \divide\epsftmp 2
       \ifnum\epsftmp>0
          \ifnum\epsftsize<\epsfrsize\else
             \advance\epsftsize-\epsfrsize \advance\epsfxsize\epsftmp \fi
       \repeat
     \fi
   \else\epsftmp=\epsfrsize \divide\epsftmp\epsftsize
     \epsfysize=\epsfxsize \multiply\epsfysize\epsftmp   
     \multiply\epsftmp\epsftsize \advance\epsfrsize-\epsftmp
     \epsftmp=\epsfxsize
     \loop \advance\epsfrsize\epsfrsize \divide\epsftmp 2
     \ifnum\epsftmp>0
        \ifnum\epsfrsize<\epsftsize\else
           \advance\epsfrsize-\epsftsize \advance\epsfysize\epsftmp \fi
     \repeat     
   \fi
   \ifepsfverbose\message{#1: width=\the\epsfxsize, height=\the\epsfysize}\fi
   \epsftmp=10\epsfxsize \divide\epsftmp\pspoints
   \vbox to\epsfysize{\vfil\hbox to\epsfxsize{%
      \includegraphics{#1}%
      \hfil}}%
\epsfxsize=0pt\epsfysize=0pt}%

{\catcode`\%=12 \global\let\epsfpercent=
\long\def\epsfaux#1#2:#3\\{\ifx#1\epsfpercent
   \def\testit{#2}\ifx\testit\epsfbblit
      \epsfgrab #3 . . . \\%
      \epsffileokfalse
      \global\epsfbbfoundtrue
   \fi\else\ifx#1\par\else\epsffileokfalse\fi\fi}%
\def\epsfgrab #1 #2 #3 #4 #5\\{%
   \global\def\epsfllx{#1}\ifx\epsfllx\empty
      \epsfgrab #2 #3 #4 #5 .\\\else
   \global\def\epsflly{#2}%
   \global\def\epsfurx{#3}\global\def\epsfury{#4}\fi}%
\def\epsfsize#1#2{\epsfxsize}
\def\eps@scaling{.95}
\def\epsscale#1{\gdef\eps@scaling{#1}}
\def\plotone#1{{\centering \leavevmode
    \epsfxsize=\eps@scaling\columnwidth \hfil \hbox{\epsfbox{#1}}}}

\newcommand{\lsun}{\hbox{{\it L}$_\odot$}\ }

\newcommand{\lstar}{L$^*$}

\newcommand{\arcsec}       {\char'175\ }

\setcounter{tocdepth}{1}

\kluwerbib

\begin{document}



\articletitle[Imaging 
of a Complete Sample of IR-Excess PG QSOs]{Imaging 
of a Complete Sample of IR-Excess PG QSOs}

\chaptitlerunninghead{QSO Hosts, Granada, 2001 }

\author{Jason A. Surace (SIRTF Science Center/Caltech) \\
and D.B. Sanders (Institute for Astronomy, University of Hawaii)}





\section{Background}

Sanders et al. (1988) proposed an evolutionary connection between Ultraluminous
Infrared Galaxies (ULIGs) and optically-selected QSOs.
In this scenario, mergers of dust and gas-rich galaxies
 provide the fuel to create and/or fuel an AGN and circumnuclear starburst. This
 dust completely enshrouds the AGN, and subsequent reradiation of the
 short-wavelength AGN emission produces the high far-IR luminositiy that defines
 ULIGs as a class. Dust clearing by superwinds eventually begins to unveil the
 central AGN, which is then perceived as a QSO. Surace et al. (1998, 2000) carried out a 
 comprehensive program of multi-wavelength high spatial resolution 
 observations using {\it HST} and ground-based tip/tilt in 
 order to characterize the morphology and colors of ULIGs. Particular 
 emphasis was given to ULIGs with ``warm'' mid-IR colors, whose 
 spectral energy distributions (SEDs) 
 and emission line features suggested that they were the most evolved 
 ULIGs in the process of becoming QSOs. These studies showed that 
 ULIGs were the merger of two \lstar\ galaxies, and that they had compact central sources 
 whose luminosity and colors are similar
 to reddened AGN. They also possess ``knots'' of star formation distributed in the
 circumnuclear regions and along the merger-generated tidal debris.

 \section{Infrared-Excess QSOs}
 
 During the evolutionary process, the SEDs of these mergers must evolve from the ULIG far-IR dominated 
 spectra towards those with a strong relative optical/UV component.
 Therefore, QSOs 
 with strong contributions to their total luminosity from far-infrared 
 emission should be less evolved and more similar to the ULIGs than 
 QSOs in general.
 
 A complete sample of 18 QSOs was selected
 from the Palomar-Green Bright Quasar Survey (Schmidt \& Green 1983), which at the time had 
 the most complete published infrared data. They lie at the same 
 distances (z$<$0.16)
 as previous samples of ULIGs examined by Surace et al., thus alleviating resolution
 dependencies in interpreting the data. ULIGs are defined to have the same minimum
 bolometric luminosity (10$^{12}$\lsun ) 
 as QSOs, as defined by Schmidt. 
 Far-infrared data from IRAS
 was used to evaluate the contribution to the bolometric luminosity of
 the ``big blue bump'' (0.1---1 $\mu$m; L$_{BBB}$) relative to that emitted in the 
 far-IR at 8---1000 $\mu$m (L$_{IR}$). All the QSOs with far-IR excesses 
 (L$_{IR}$/L$_{BBB}$) as great as the least far-IR active
 ULIG (3C 273; L$_{IR}$/L$_{BBB}$=0.46) were selected. A campaign of high resolution observations at B, I, H,
 and K' using a fast tip/tilt guider on the UH 2.2m telescope was 
 carried out on 17 out of the complete sample
 of 18 objects. Typical spatial resolutions at H and K' were 0.25\arcsec 
 while those at B and I were
 0.7\arcsec. The data were either photometric or were tied to that
 of Neugebauer et al. (1987) so that colors could be derived.
 
 \begin{figure}[tbh]
 \epsscale{0.63}
 \plotone{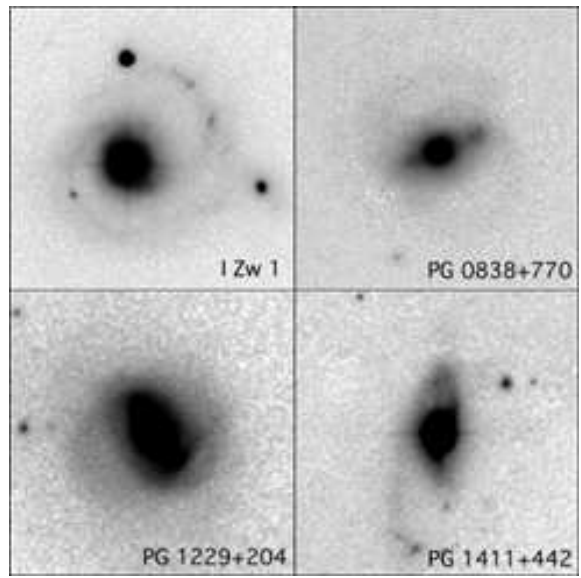}
 \caption{Optical images showing the diversity in host galaxy 
 morphology of Infrared-Excess PG QSOs, including a spiral galaxy, two 
 barred spirals, and a major merger with an 80 kpc tidal tail.}
 \end{figure}
 
\section{Results}
 
1) All of the IR-excess QSOs have readily detectable hosts.
Their morphologies are varied, but at least 50\% are 
spiral-type systems, as evidenced by the presence of spiral arms,
and just over half of these are barred. This is similar to 
results for Seyfert galaxies and radio quiet quasars (McLeod \& Rieke 
1994/95; Taylor et al. 1996). The variance between this 
result and other recent imaging studies (McLure et al. 1999), 
which have found a predominance of elliptical-like hosts, is probably 
due to selection effects. Our infrared criterion selects galaxies 
with significant reservoirs of gas and dust (Evans et al., this 
vol.) and is known from {\it 
IRAS} to strongly select spiral galaxies. Furthermore, the sample does not 
match the luminosity {\it distribution} of the Surace et al. ULIGs. 
Instead, the demand for low redshift systems selects the 
least-luminous QSOs, as opposed to the higher-luminosity samples used 
by others. 

2) The mean H-band luminosity of the (point-source subtracted) host 
galaxies is 2.2 \lstar. The distribution of host galaxy
 luminosities for ``warm'' ULIGs and QSOs is nearly identical.

3) 22\% are in major merger systems. An additional 25\% have 
barred morphologies that are consistent with late-stage minor merger 
morphologies. The total number of host galaxies where mergers may be 
implicated is thus between 25---50\%. The remaining 25\% 
with indeterminate or elliptical-like features may also be merger 
remnants.

4) Several have small-scale ($<$1 kpc) structure similar to that
 seen in ULIGs; these knots are typically very red and are similar to 
 dust-enshrouded star formation with an age of $\leq$ 100 Myrs.
 The QSO nuclei also have near-IR excesses (beyond dust-reddening)
 which may be the result of small amounts of hot thermal dust emission.
  
5) These results imply that while some 25\% of IR-excess QSOs have morphologies
 consistent with advanced mergers, another 
25\% (apparently unperturbed spiral galaxies) cannot have had their infrared 
activity triggered via the major mergers implicated in the formation of ULIGs. 
A more thorough understanding of these 
statistics will necessitate similar obervations 
of the complementary sample of non-IR-excess QSOs. A larger complete sample of 
QSOs with greater counting statistics would also be useful.
 





%




\end{document}